\documentclass[sn-mathphys,Numbered]{sn-jnl}

\usepackage{graphicx}%
\usepackage{multirow}%
\usepackage{amsmath,amssymb,amsfonts}%
\usepackage{amsthm}%
\usepackage{mathrsfs}%
\usepackage[title]{appendix}%
\usepackage{xcolor}%
\usepackage{textcomp}%
\usepackage{manyfoot}%
\usepackage{booktabs}%
\usepackage{algorithm}%
\usepackage{algorithmicx}%
\usepackage{algpseudocode}%
\usepackage{listings}%

\usepackage[T1]{fontenc}
\usepackage{frcursive}
\usepackage{calligra}

\theoremstyle{thmstyleone}%
\theoremstyle{thmstyletwo}%

\theoremstyle{thmstylethree}%

\begin{document}

\title[Gauging Centrifugal Instabilities in Compressible Free-Shear Layers Via Nonlinear Boundary Region Equations]{Gauging Centrifugal Instabilities in Compressible Free-Shear Layers Via Nonlinear Boundary Region Equations}


\author*[1]{\fnm{Omar} \sur{Es-Sahli}}

\author[1]{\fnm{Adrian} \sur{Sescu}}

\author[2]{\fnm{Yuji} \sur{Hattori}}

\affil*[1]{\orgdiv{Department of Aerospace Engineering}, \orgname{Mississippi State University}, \orgaddress{\street{75 B. S. Hood Rd}, \city{Mississippi State}, \postcode{39762}, \state{MS}, \country{USA}}}

\affil[2]{\orgdiv{Institute of Fluid Science}, \orgname{Tohoku University}, \orgaddress{\street{2 Chome-1-1 Katahira}, \city{Sendai}, \postcode{980-8577}, \country{Japan}}}


\abstract{Curved free shear layers emerge in many engineering problems involving complex flow geometries, such as the flow over a backward facing step, flows with wall injection in a boundary layer, the flow inside side-dump combustors, or wakes generated by vertical axis wind turbines, among others. Previous studies involving centrifugal instabilities have mainly focused on wall-flows where Taylor instabilities between two rotating concentric cylinders or G\"{o}rtler vortices in boundary layers, resulting from the imbalance between the centrifugal forces and the radial pressure gradients, are generated.
Curved free shear layer flows, however, have not received sufficient attention, especially in the nonlinear regime. 
The present work investigates the development of centrifugal instabilities evolving in a curved free shear layer flow in the nonlinear compressible regime.
The compressible Navier-Stokes equations are reduced to the nonlinear boundary region equations (BRE) in a high Reynolds number asymptotic framework wherein the streamwise wavelengths of the disturbances are assumed to be much larger than the spanwise and wall-normal counterparts.
We study the effect of the freestream Mach number $\boldsymbol{M_\infty}$, the shear layer thickness $\boldsymbol{\delta}$, the amplitude of the incoming disturbance $\boldsymbol{A}$, and the relative velocity difference across the shear layer $\boldsymbol{\Delta V}$ on the development of these centrifugal instabilities. 
Our parametric study shows, among other things, that the kinetic energy of the curved shear layer flow is directly proportional to $\boldsymbol{\Delta V}$ and $\boldsymbol{A}$ while it is inversely proportional to $\boldsymbol{\delta}$.}

\maketitle

\section{Introduction}

The stability of curved shear layer flows depends on the velocity difference across the shear layer and the radius of the curvature. For a free shear layer with no curvature, as known as a plane shear layer, the Kelvin-Helmholtz instability is the dominant turbulent mechanism \cite{Michalke}. In this case, the two-dimensional disturbances are more unstable than their three-dimensional counterpart \cite{Squire}, and the Kelvin-Helmholtz instability introduces predominantly spanwise oriented vortices. Rayleigh \cite{Rayleigh} proved that the presence
of an inflection point in the basic velocity profile is a necessary condition for the Kelvin-Helmholtz instability. On the other hand, for a curved mixing layer flow with an inflectional velocity profile, the Kelvin-Helmholtz instability mechanism is still present, along with centrifugal instabilities in the form of streamwise oriented G\"{o}rtler type vortices.

G\"{o}rtler vortices are mostly known to appear inside a boundary layer flow along a concave surface due to the imbalance between radial pressure gradients and centrifugal forces  (e.g., Gortler \cite{Gortler}, Hall \cite{hall1}, Swearingen \& Blackwelder \cite{Swearingen}). For highly curved walls, for example, vortex formation occurs more rapidly and can significantly alter the mean flow causing the laminar flow to transition into turbulence. Under certain conditions, G\"{o}rtler vortices can be efficient precursors to transition. The growth rate of these counter-rotating streamwise vortical structures depends on the surface curvature and the receptivity of the boundary layer to freestream disturbances and surface imperfections. This type of instability is predominant in many engineering applications. Thus, understanding the processes leading to their development and predicting their occurrence using efficient and tractable methods could potentially advance the overall understanding of turbulence.

Numerous theoretical and numerical studies covered centrifugal instabilities in incompressible curved free shear layer flows. Plesniak et al. \cite{Plesniak1, Plesniak2} conducted extensive experimental measurements investigating curved two-stream mixing layers to show how centrifugal effects yield streamwise vortices. The untripped case within this suite of experiments exhibited organized streamwise vorticity, while the tripped case did not. The researchers explain that this is due to the spatially stationary streamwise vortices that provide extra entrainment to the flow in the tripped case (Bell \& Mehta \cite{Bell} showed this for a plane mixing layer). Hu et al. \cite{Hu} and Liou \cite{Liou} focused on the effect of the curvature on the inflectional Rayleigh modes, which they found to be minimal, although the curvature excites an unstable three-dimensional disturbance with the amplitude increasing as the streamwise wavenumber decreases. The analytical and numerical study of Otto et al. \cite{Otto} showed that the unstable modes largely depend on surface curvature. They also employed numerical simulations to solve the parabolic equations, assuming that the wavenumber and G\"{o}rtler number are both of order one. They found that as the difference between the freestream speeds increased, the layer became more susceptible to centrifugal instabilities.

In the present work, we analyze the development of centrifugal instabilities in high-speed compressible curved free shear layer flows via an efficient numerical algorithm based on the nonlinear boundary region equations (NBREs) -- a parabolized version of the Navier-Stokes equations under the assumption that the streamwise wavenumber associated with the disturbances is much smaller than the cross-flow wavenumbers. The study considers the effects of a wide range of Mach numbers, the amplitude of the freestream disturbance, $A$, the shear layer thickness, $\delta$, and the velocity difference across the shear layer, $\Delta V$, on the development and growth of these centrifugal instabilities. The study shows, among other things, that the kinetic energy level of the curved shear layer flow is directly proportional to $\Delta V$ and $A$ while it is inversely proportional to $\delta$. Increasing $A$ induces larger instability structures, as expected, which may be beneficial for enhancing mixing. The location of the maximum energy moves farther downstream as the freestream Mach number increases.

\section{Problem formulation and numerical algorithm}

\subsection{Scalings}


All dimensional spatial coordinates $(x^*,y^*,z^*)$ are normalized by the spanwise separation $\lambda^*$, while the dependent variables by their respective freestream values, except the pressure, which is normalized by the dynamic pressure:

\begin{eqnarray}
\bar{t} = \frac{t^*}{\lambda^*/V_{\infty}^*}; \hspace{4mm} \bar{x} = \frac{x^*}{\lambda^*}; \hspace{4mm}
\bar{y} = \frac{y^*}{\lambda^*}; \hspace{4mm} \bar{z} = \frac{z^*}{\lambda^*}
\end{eqnarray}

\begin{eqnarray}
\bar{u} = \frac{u^*}{V_{\infty}^*}; \hspace{4mm} \bar{v} = \frac{v^*}{V_{\infty}^*}; \hspace{4mm}
\bar{w} = \frac{w^*}{V_{\infty}^*}; \hspace{4mm} \bar{\rho} = \frac{\rho^*}{\rho_{\infty}^*}
\end{eqnarray}

\begin{eqnarray}
\bar{p} =  \frac{p^*-p_{\infty}^*}{\rho_{\infty}^*V_{\infty}^{*2}}; \hspace{4mm} 
\bar{T} = \frac{T^*}{T_{\infty}^*}; 
\hspace{4mm} \bar{\mu} = \frac{\mu^*}{\mu_{\infty}^*}; \hspace{4mm} \bar{k} = \frac{k^*}{k_{\infty}^*}
\end{eqnarray}
where $\lambda^*$ is the spanwise wavelength of the disturbances, $(u^*,v^*,w^*)$ are the velocity components, $\rho^*$ the density, $p^*$ is pressure, $T^*$ temperature, $\mu^*$ dynamic viscosity, $k^*$ thermal conductivity, and all quantities with $\infty$ at the subscript represent conditions at infinity.

Reynolds number based on the spanwise separation, Mach number and Prandtl number are defined as

\begin{eqnarray}
R_{\lambda} = \frac{\rho_{\infty}^* V_{\infty}^* \lambda^*}{\mu_{\infty}^*}, \hspace{5mm}
M_{\infty} = \frac{V_{\infty}^*}{a_{\infty}^*}, \hspace{5mm}
Pr = \frac{\mu_{\infty}^* C_p}{k_{\infty}^*}
\end{eqnarray}
where $\mu_{\infty}^*$, $a_{\infty}^*$ and $k_{\infty}^*$ are freestream dynamic viscosity, speed of sound and thermal conductivity, respectively, and $C_p$ is the specific heat at constant pressure. As for boundary layer flows over curved surfaces, we here define the equivalent global G\"{o}rtler number as

\begin{eqnarray}
G_{\lambda} = \frac{R_{\lambda}^2 \lambda^*}{r^*}
\end{eqnarray}
where $r^*$ is the radius of the curvature.

\subsection{Boundary region equations: a parabolized form of the Navier-Stokes equations}

If the streamwise wavenumber of the disturbances evolving inside the shear layer are much larger that the wavenumbers corresponding to the crossflow directions, then the Navier-Stokes equations can be transformed into a parabolic set of equations in the framework of high Reynolds number asymptotics. 

For a full compressible, Newtonian flow, the primitive form of the Navier-Stokes equations with non-dimensional variables are considered here in the form 
 
 \small

\begin{eqnarray}
\frac{D \bar{\rho}}{D t} 
+ \rho \left( \frac{\partial \bar{u}}{\partial \bar{x}} + \frac{\partial \bar{v}}{\partial \bar{y}} + \frac{\partial \bar{w}}{\partial \bar{z}} \right) = 0
\end{eqnarray}

\begin{eqnarray}
\bar{\rho} \frac{D \bar{u}}{D \bar{t}}
= -\frac{\partial \bar{p}}{\partial \bar{x}} 
+ \frac{1}{Re_{\lambda}}\frac{\partial}{\partial \bar{x}} \left[ \frac{2}{3} \mu \left( 2\frac{\partial \bar{u}}{\partial \bar{x}} - \frac{\partial \bar{v}}{\partial \bar{y}} - \frac{\partial \bar{w}}{\partial \bar{z}} \right) \right]
+ \frac{\partial}{\partial \bar{y}} \left[ \mu \left( \frac{\partial \bar{u}}{\partial \bar{y}} + \frac{\partial \bar{v}}{\partial \bar{x}} \right) \right]
+ \frac{\partial}{\partial \bar{z}} \left[ \mu \left( \frac{\partial \bar{w}}{\partial \bar{x}} + \frac{\partial \bar{u}}{\partial \bar{z}} \right) \right]
\end{eqnarray}

\begin{eqnarray}
\bar{\rho} \frac{D \bar{v}}{D \bar{t}}
= -\frac{\partial \bar{p}}{\partial \bar{y}} 
+ \frac{1}{Re_{\lambda}}\frac{\partial}{\partial \bar{y}} \left[ \frac{2}{3} \mu \left( 2\frac{\partial \bar{v}}{\partial \bar{y}} - \frac{\partial \bar{u}}{\partial \bar{x}} - \frac{\partial \bar{w}}{\partial \bar{z}} \right) \right]
+ \frac{\partial}{\partial \bar{x}} \left[ \mu \left( \frac{\partial \bar{v}}{\partial \bar{x}} + \frac{\partial \bar{u}}{\partial \bar{y}} \right) \right]
+ \frac{\partial}{\partial \bar{z}} \left[ \mu \left( \frac{\partial \bar{v}}{\partial \bar{z}} + \frac{\partial \bar{w}}{\partial \bar{y}} \right) \right] 
\end{eqnarray}

\begin{eqnarray}
\bar{\rho} \frac{D \bar{w}}{D \bar{t}}
= -\frac{\partial \bar{p}}{\partial \bar{z}} 
+ \frac{1}{Re_{\lambda}}\frac{\partial}{\partial \bar{z}} \left[ \frac{2}{3} \mu \left( 2\frac{\partial \bar{w}}{\partial \bar{z}} - \frac{\partial \bar{u}}{\partial \bar{x}} - \frac{\partial \bar{v}}{\partial \bar{y}} \right) \right]
+ \frac{\partial}{\partial \bar{x}} \left[ \mu \left( \frac{\partial \bar{w}}{\partial \bar{x}} + \frac{\partial \bar{u}}{\partial \bar{z}} \right) \right]
+ \frac{\partial}{\partial \bar{y}} \left[ \mu \left( \frac{\partial \bar{v}}{\partial \bar{z}} + \frac{\partial \bar{w}}{\partial \bar{y}} \right) \right]
\end{eqnarray}

\begin{eqnarray}
&& \bar{\rho} \frac{D \bar{T}}{D \bar{t}}
= 
\frac{1}{Pr Re_{\lambda}} \left[ \frac{\partial}{\partial \bar{x}} \left( k \frac{\partial \bar{T}}{\partial \bar{x}} \right) + \frac{\partial}{\partial \bar{y}} \left( k \frac{\partial \bar{T}}{\partial \bar{y}} \right) + \frac{\partial}{\partial \bar{z}} \left( k \frac{\partial \bar{T}}{\partial \bar{z}} \right) \right]  \nonumber  \\
&-& (\gamma - 1) M_{\infty}^2 \left[ p \left( \frac{\partial \bar{u}}{\partial \bar{x}} + \frac{\partial \bar{v}}{\partial \bar{y}} + \frac{\partial \bar{w}}{\partial \bar{z}} \right)
-\frac{2}{3} \mu \left( \frac{\partial \bar{u}}{\partial \bar{x}} + \frac{\partial \bar{v}}{\partial \bar{y}} + \frac{\partial \bar{w}}{\partial \bar{z}} \right)^2 \right] \\
&+& (\gamma - 1) M_{\infty}^2 \frac{\mu}{Re_{\lambda}} \left[ 2\left( \frac{\partial \bar{u}}{\partial \bar{x}} \right)^2 + 2\left( \frac{\partial \bar{v}}{\partial \bar{y}} \right)^2 + 2\left( \frac{\partial \bar{w}}{\partial \bar{z}} \right)^2
+ \left( \frac{\partial \bar{u}}{\partial \bar{y}} + \frac{\partial \bar{v}}{\partial \bar{x}} \right)^2
+ \left( \frac{\partial \bar{w}}{\partial \bar{x}} + \frac{\partial \bar{u}}{\partial \bar{z}} \right)^2
+ \left( \frac{\partial \bar{v}}{\partial \bar{z}} + \frac{\partial \bar{w}}{\partial \bar{y}} \right)^2 \right]   \nonumber
\end{eqnarray}
\normalsize
where 

\begin{eqnarray}
\frac{D}{D\bar{t}} = \frac{\partial}{\partial \bar{t}} + \bar{u} \frac{\partial}{\partial \bar{x}} + \bar{v} \frac{\partial}{\partial \bar{y}} + \bar{w} \frac{\partial}{\partial \bar{z}}
\end{eqnarray}
is the substantial derivative. The pressure $\bar{p}$, the temperature $\bar{T}$  and the density $\bar{\rho}$ of the fluid are combined in the equation of state in non-dimensional form, $\bar{p} = \bar{\rho} \bar{T} / \gamma M_{\infty}^2$, assuming that non-chemically-reacting flows are considered. Other notations include the dynamic viscosity $\mu$, and the free-stream Mach number $M_{\infty}=V_{\infty}^*/a_{\infty}^*$. The dynamic viscosity $\mu$ and thermal conductivity $k$ are linked to the temperature using a power law in dimensionless form,

\begin{eqnarray}
\mu = T^b;  \hspace{6mm}
k = \frac{C_p \mu}{Pr}
\end{eqnarray}
where $b=0.76$ (Ricco \& Wu \cite{Ricco1}), $C_p = \gamma R / (\gamma - 1)$, $\gamma = 1.4$, and $Pr = 0.72$ for air.

We re-scale the streamwise distance and time co-ordinate at which the vortex system forms by the following $\mathcal{O}(1)$ variables: $x = \bar{x}/R_{\lambda}$, and the time as $t = \bar{t}/R_{\lambda}$. Note that the distance in the wall-normal and spanwise directions are the same, $y = \bar{y}$, $z = \bar{z}$. Another thing to mention is that, in this region, the crossflow velocity component is small compared to the streamwise velocity component, and pressure variations are negligible. Appropriate dominant balance considerations suggest that the dependent variables in this region must also re-scale as follows:

\begin{eqnarray}
u = \bar{u}; \hspace{4mm}  v = \bar{v} /R_{\lambda}; \hspace{4mm}  w = \bar{w} /R_{\lambda}; \hspace{4mm}  
\rho = \bar{\rho}; \nonumber  \\
p = \bar{p} /R_{\lambda}^2; \hspace{4mm} T = \bar{T}; \hspace{4mm}  
\mu = \bar{\mu}; \hspace{4mm}  k = \bar{k};
\end{eqnarray}
Working out the order-of-magnitude analysis of the Navier-Stokes equations, we obtain the parabolic set of equations, which we refer to as the nonlinear compressible boundary region equations (NCBRE) 
\small
	\begin{equation}\label{neq1}
		\vec{V} \cdot \nabla \rho 
		+ \rho \nabla \cdot \vec{V} = 0
	\end{equation}
	
	\begin{equation}\label{neq2}
		\rho \vec{V} \cdot \nabla u
		= 
		\nabla_{c} \cdot \left( \mu \nabla_{c} u \right)
	\end{equation}
	
	\begin{equation}\label{neq3}
		\rho \vec{V} \cdot \nabla v + G_\lambda u^2
		= 
		-\frac{\partial p}{\partial y} 
		+ \frac{\partial}{\partial y} \left[ \frac{2}{3} \mu \left( 3\frac{\partial v}{\partial y} - \nabla \cdot \vec{V} \right) \right] 
		+ \frac{\partial}{\partial x} \left( \mu \frac{\partial u}{\partial y} \right) 
		+ \frac{\partial}{\partial z} \left[ \mu \left( \frac{\partial v}{\partial z} + \frac{\partial w}{\partial y} \right) \right]
	\end{equation}
	
	\begin{equation}\label{neq4}
		\rho \vec{V} \cdot \nabla w
		= 
		-\frac{\partial p}{\partial z} 
		+ \frac{\partial}{\partial z} \left[ \frac{2}{3} \mu \left( 3\frac{\partial w}{\partial z} - \nabla \cdot \vec{V} \right) \right]
		+ \frac{\partial}{\partial x} \left( \mu \frac{\partial u}{\partial z} \right)
		+ \frac{\partial}{\partial y} \left[ \mu \left( \frac{\partial v}{\partial z} + \frac{\partial w}{\partial y} \right) \right]
	\end{equation}
	
	\begin{equation}\label{neq5}
		\rho \vec{V} \cdot \nabla T
		= 
		\frac{1}{Pr} \nabla_{c} \cdot \left( k \nabla_{c} T \right)
		+(\gamma - 1) M_{\infty}^2 \mu \left[ \left( \frac{\partial u}{\partial y} \right)^2
		+ \left( \frac{\partial u}{\partial z} \right)^2 \right]
	\end{equation}
	\normalsize
where $\vec{V}$ is the velocity vector and $\nabla_{c}$ is the crossflow nabla operator:
	
	\begin{equation}
	\vec{V} = u \vec{i} + v \vec{j} + w \vec{k}; \hspace{6mm} \nabla_{c} = \frac{\partial}{\partial y} \vec{j} + \frac{\partial}{\partial z} \vec{k} \end{equation}
	The effect of the wall curvature is contained in the term involving the global G\"{o}rtler number $G_\lambda$ in the second momentum equation. \\

A small artificial disturbance is imposed at the inflow boundary in the form:

\begin{equation}\label{disturbance}
v' = A \cos \left( \frac{\pi z}{\lambda} \right) \exp \left[ -\frac{(y-y_0)^2}{\sigma^2} \right]
\end{equation}
where $A$ is a small amplitude (in this study $A = 0.04$). $\lambda^*$ is the spanwise wavenumber (dictating the spanwise separation of the centrifugal instabilities), and $\sigma$ represents the extent of the disturbance in the $y$ direction. In the present work, $\lambda$ is kept constant at 0.8 cm. Es-Sahli et al. \cite{EsSahli1} elaborately studied the effect of $\lambda^*$ on the development of centrifugal instabilities in curved free shear layers. The NCBREs are solved using the algorithm developed in Es-Sahli et al. \cite{Es-Sahli2} and previously in Sescu and Thompson \cite{Sescu3}. We slightly adjusted the algorithm to accommodate the free shear layer setting.

\section{Results}

In this section, we present and discuss a set of results from the curved free shear layer numerical simulations. The flow domain is split into `fast' and `slow' streams, both having velocities $V_f$ and $V_s$, respectively, which we define as; $V_f = V_\infty$ and $V_s = (1-\Delta V)V_\infty$, where $\Delta V$ is the relative velocity difference (here, we set this difference at four levels, $20\%$, $30\%$, $40\%$, and $50\%$). The temperature in the fast stream is set to $T_\infty$, while in the slow stream is set to $0.9T_\infty$. In our parametric study, we consider three Mach numbers in the fast stream of $2.0$, $4.0$, and $6.0$, respectively (the lowest velocity in the slow stream will correspond to a Mach number of $1.0$). We also consider three values for the shear layer thickness $\delta$ at the inflow boundary at $0.2$, $0.4$, and $0.6$, where the velocity variation between the fast and slow streams is modeled via a hyperbolic tangent function $0.5(1-\tanh{(y-y_0)/\delta})$, whith $y_0$ representing the location of the shear layer. A similar function is used to model the variation of the temperature in the shear layer, with the same thickness, although in reality the thickness of the thermal layer may be slightly different. 

The Reynolds number $R_{\lambda}$ based on the faster freestream and the spanwise separation of the disturbance and the global G\"{o}rtler number $G_{\lambda}$ are kept the same for all cases at $10^{6}$ and $2\times 10^5$, respectively (the kinematic viscosity and the curvature of the wall were varied to achieve constant $R_{\lambda}$ and $G_{\lambda}$ for all simulations). The grid is uniform in the spanwise direction taking into account that the flow is periodic in this direction, while in the radial direction the grid is stretched towards the top and bottom far-field boundaries. The marching in the streamwise direction is achieved by means of an explixit Euler method, with equally-spaced discretization. At the inflow boundary, centrifugal instabilities are excited by the non-dimensional artificial disturbance in equation \ref{disturbance} imposed on the base flow at the inflow boundary, with the amplitude set at $0.04$. 


\begin{figure}[htp]
 \begin{center}
 \includegraphics[width=11cm]{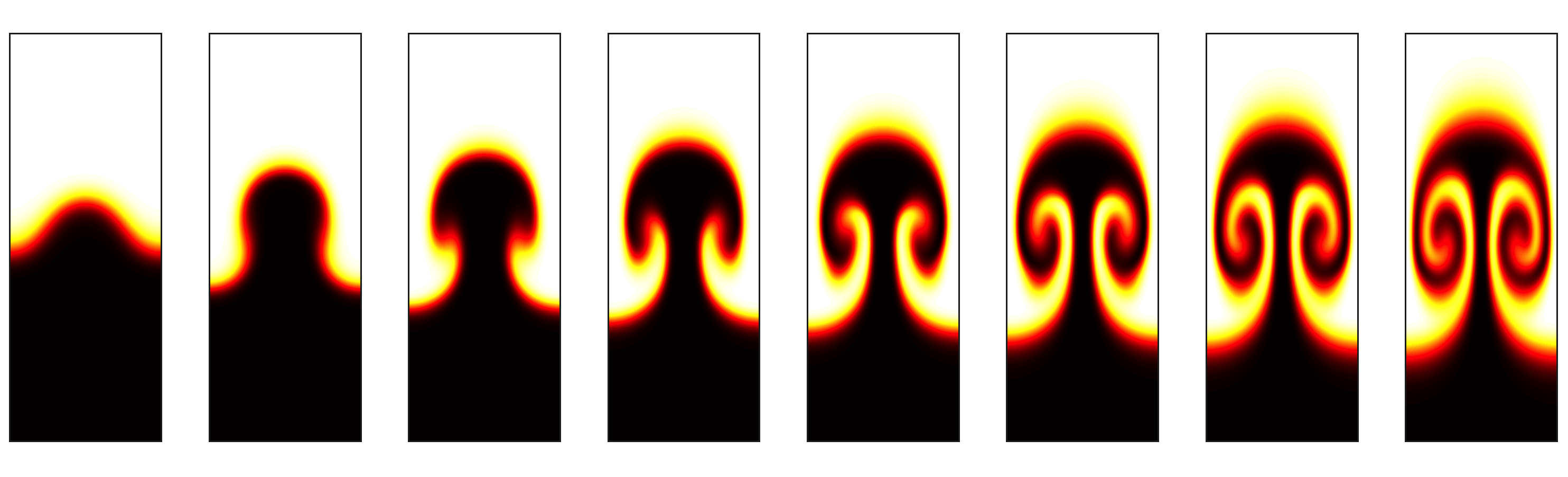}
 \includegraphics[width=11cm]{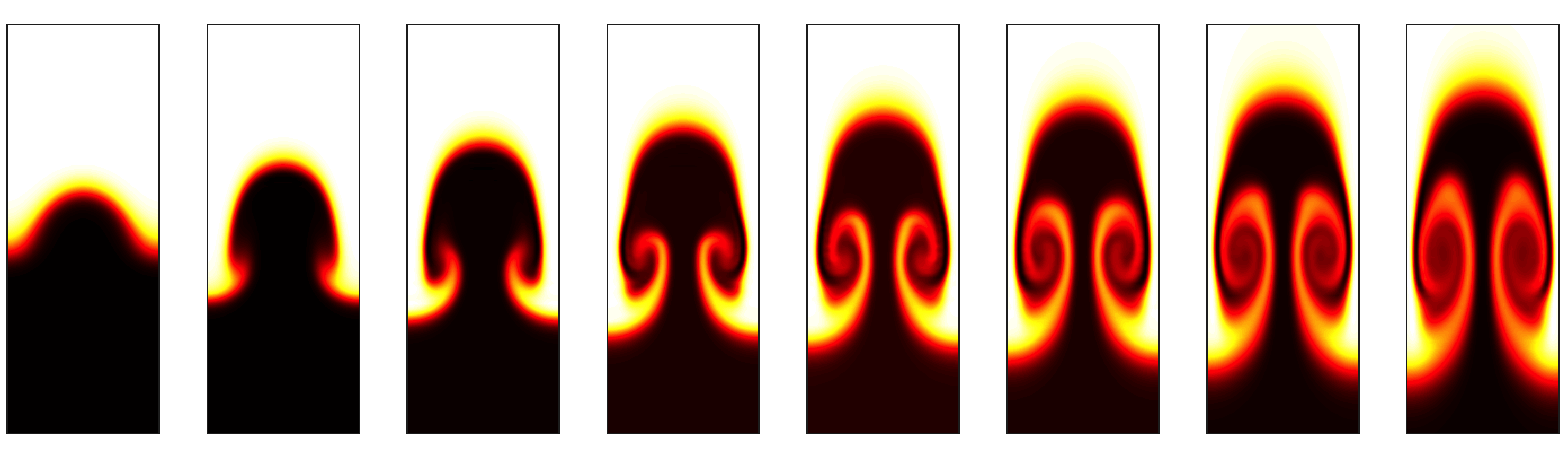}
 \end{center}
  \caption{Contour plots of the streamwise velocity $\boldsymbol{u}$ at different streamwise locations for $M=2$ and $\delta=0.2$: top - $V_s=80\% V_f$; bottom - $V_s=50\% V_f$.}
  \label{f1}
\end{figure}

\begin{figure}[htp]
 \begin{center}
 \includegraphics[width=11cm]{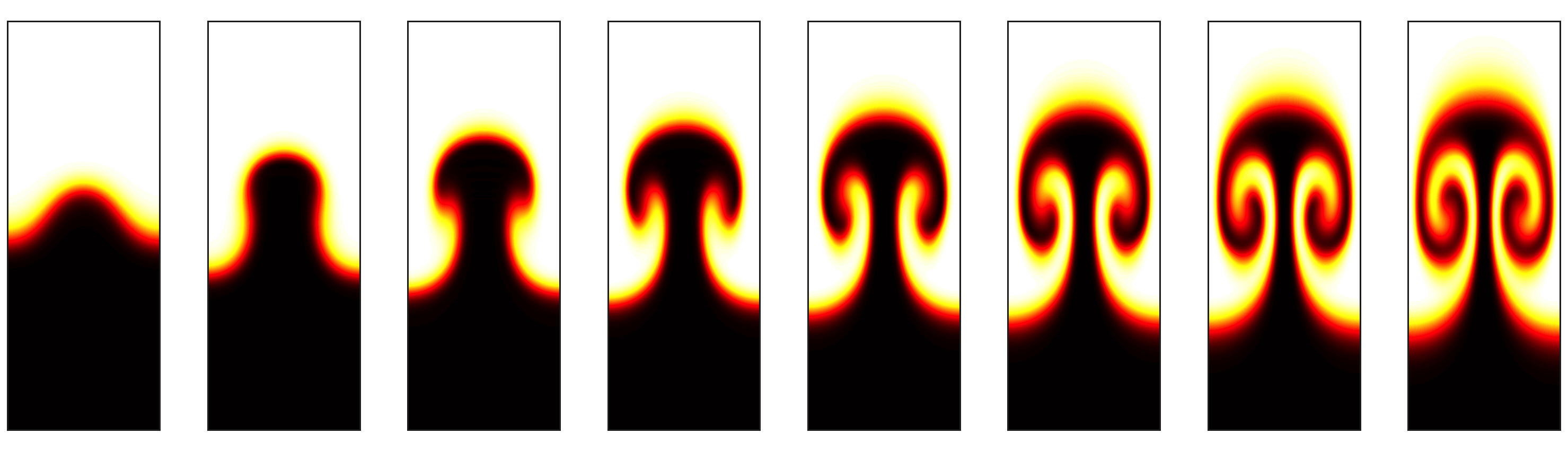}
 \includegraphics[width=11cm]{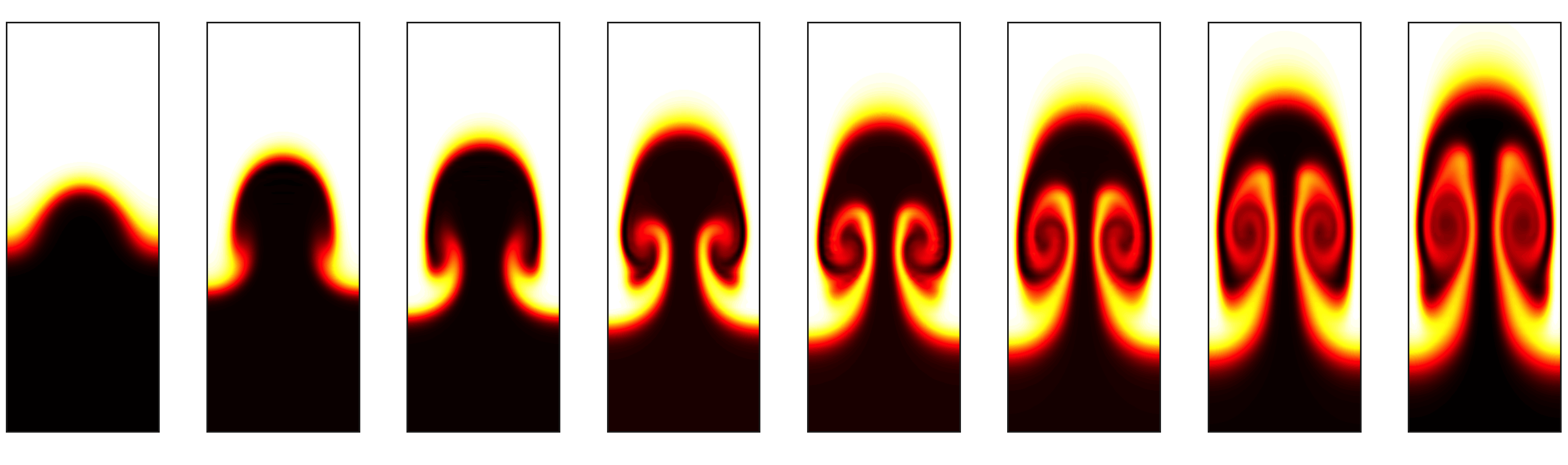}
 \end{center}
  \caption{Contour plots of the streamwise velocity $\boldsymbol{u}$ at different streamwise locations for $M=4$ and $\delta=0.2$: top - $V_s=80\% V_f$; bottom - $V_s=50\% V_f$.}
  \label{f2}
\end{figure}

\begin{figure}[htp]
 \begin{center}
 \includegraphics[width=11cm]{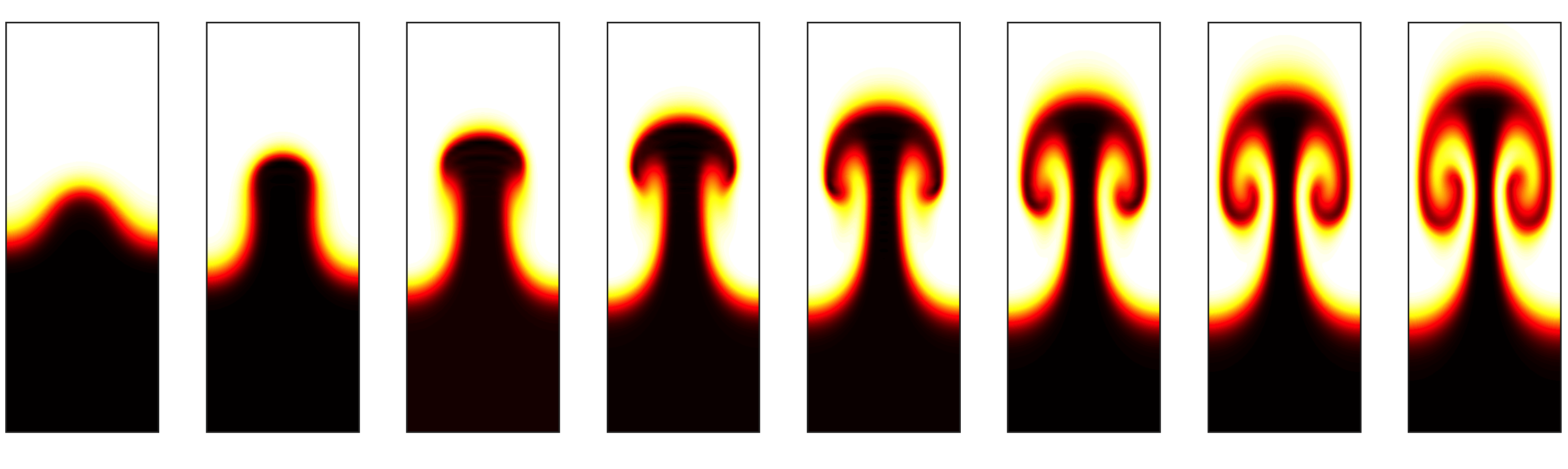}
 \includegraphics[width=11cm]{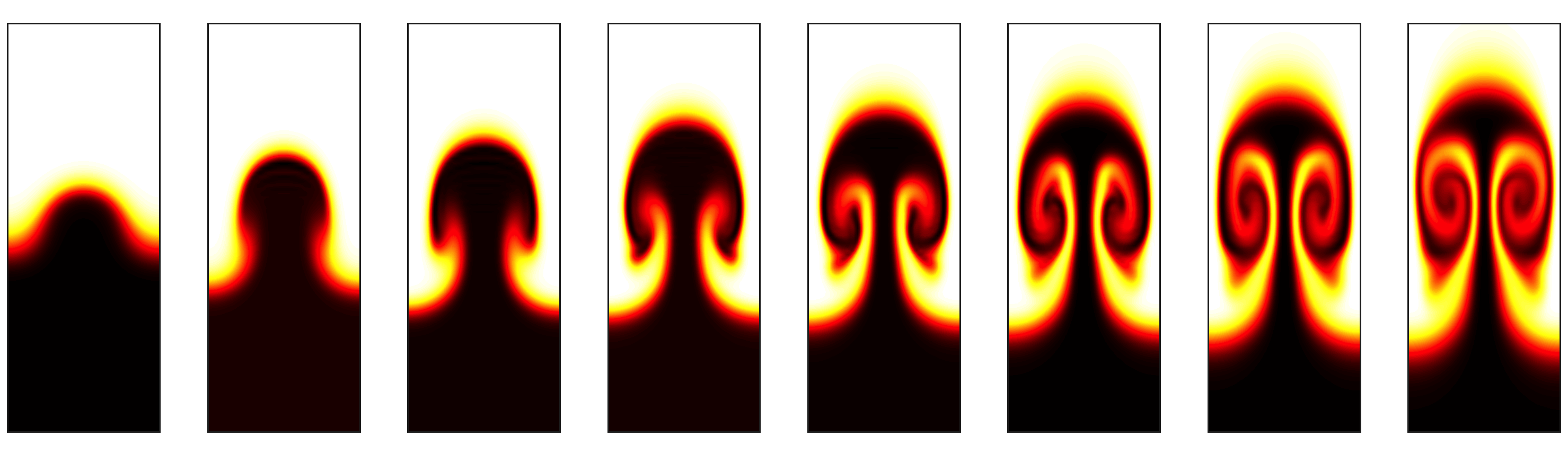}
 \end{center}
  \caption{Contour plots of the streamwise velocity $\boldsymbol{u}$ at different streamwise locations for $M=6$ and $\delta=0.2$: top - $V_s=80\% V_f$; bottom - $V_s=50\% V_f$.}
  \label{f3}
\end{figure}

Figures \ref{f1}-\ref{f3} present consecutive contour plots depicting the magnitude of crossflow velocity for different Mach numbers and velocity differences $\Delta V$, with the shear layer thickness $\delta$ set to $0.2$. The color scheme represents the flow velocity, with white at the top indicating the fast stream and black at the bottom indicating the slow stream. These contour plots illustrate the progression of the centrifugal instabilites in the streamwise direction; we hypothesize that the incoming disturbance is sufficiently high to allow the centrifugal instabilities to take over the Kelvin-Helmholtz type instabilities (this mathematical model is not capable of predicting Kelvin-Helmholtz type instabilities because the problem is steady and the system of equations is parabolic). To characterize the shape of the centrifugal instabilities, we identify two types of structures: "primary" and "secondary." The primary flow structure refers to the mushroom-like formation that evolves as the main instability, as exemplified by the fourth panel of the first row in figure \ref{f1}. On the other hand, the secondary structures are elongated features that emerge from the edges of the primary flow structure, as seen, for example, in the fourth panel of the second row in figure \ref{f2}.

By comparing the first to the second row of each figure \ref{f1}, \ref{f2} or \ref{f3}, we notice that by increasing the velocity difference across the shear layer, $\Delta V$, accelerates the development of the mushroom-like primary structures and makes the secondary structures more prominent. The color transition from black to white seen in the bottom rows of each of these figures suggest that the higher the velocity difference across the shear layer, the more intense the mixing in the shear layer is. Comparing the centrifugal instabilities at the same streamwise coordinate across various Mach numbers, we observe a delay in the growth of the mushroom-like structures as $M_\infty$ increases. We also observe that the flow structures become thinner as the Mach number increases. For instance, in the third panel of the first row in figure \ref{f1}, for which $M_\infty = 2$, the mushroom structure has already begun to form, whereas in the corresponding panel of figure \ref{f3}, where $M_\infty = 6$, the mushroom shape has not yetr developed. This finding suggests that, for high Mach number free shear layer flows, the same mixing efficiency is achieved further downstream. 

\begin{figure}[htp]
 \begin{center}
  \includegraphics[width=11cm]{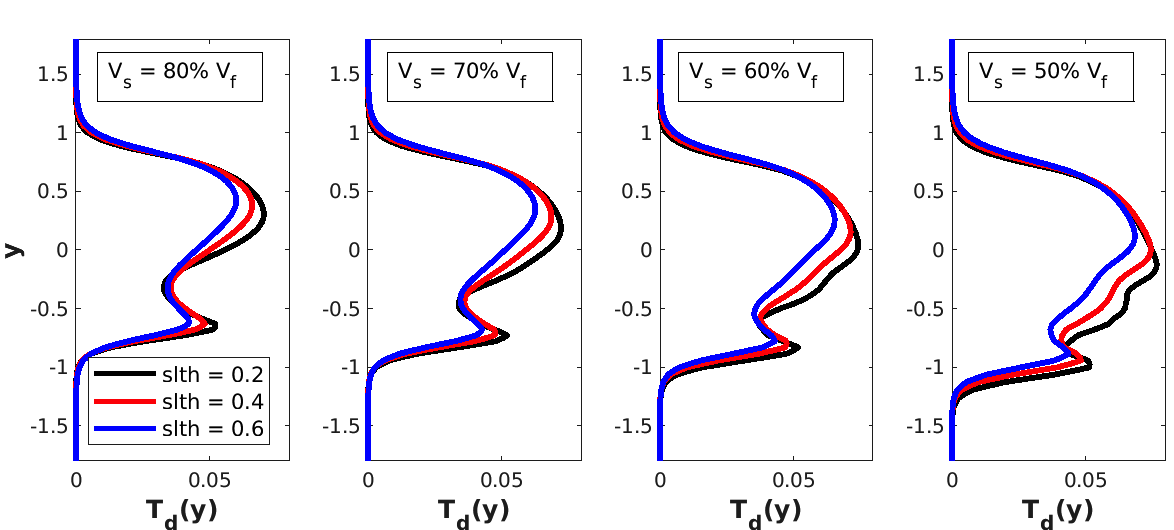}
 \end{center}
  \caption{Profiles of temperature disturbance $T_d(y)$ for the $\boldsymbol{M=2}$ case.}
  \label{f4}
\end{figure}

\begin{figure}[htp]
 \begin{center}
  \includegraphics[width=11cm]{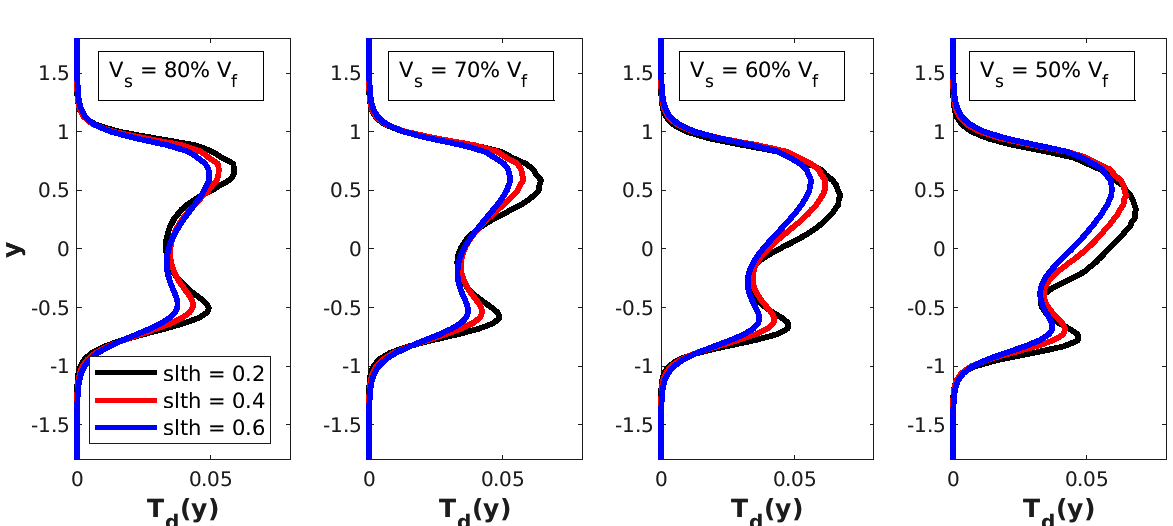}
 \end{center}
  \caption{Profiles of temperature disturbance $T_d(y)$ for the $\boldsymbol{M=4}$ case.}
  \label{f5}
\end{figure}

\begin{figure}[htp]
 \begin{center}
  \includegraphics[width=11cm]{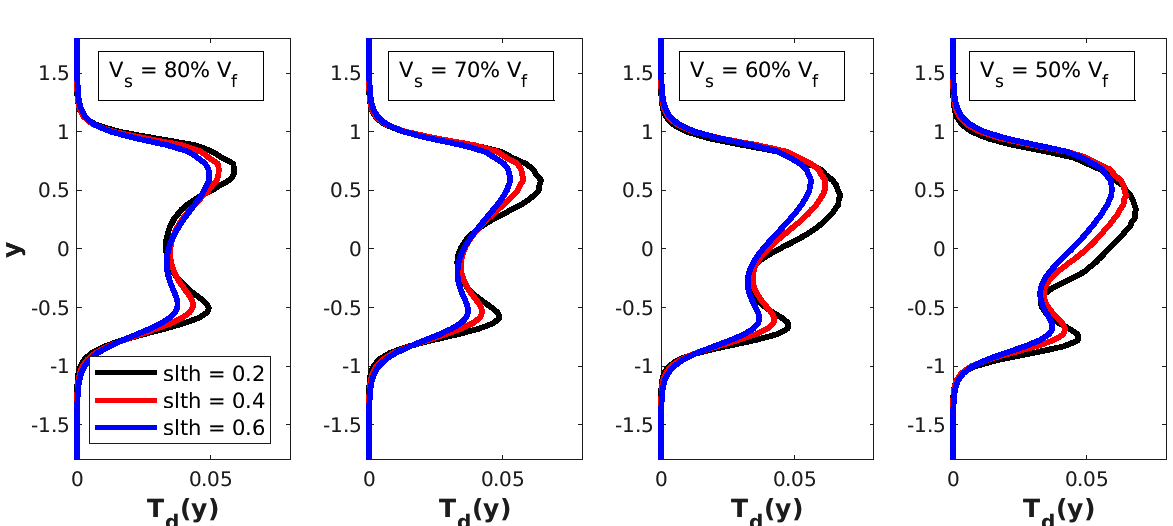}
 \end{center}
  \caption{Profiles of temperature disturbance $T_d(y)$ for the $\boldsymbol{M=6}$ case.}
  \label{f6}
\end{figure}

To quantify the thermal effects of these instabilities, vertical profiles of temperature disturbance $T_d(x,y,z) = T(x,y,z) - T_m(x,y)$, where $T_m(x,y)$ is the spanwise mean component of temperature $T(x,y,z)$, through the center of the mushroom shape are included in figures \ref{f4}-\ref{f6}. The profiles are compared to each other for different shear layer thicknesses and different velocity difference levels across the shear layer. They all show that increasing the thickness of the shear layer increases the temperature disturbance, and that this increase is slightly less significant at high Mach numbers. As expected, by increasing the velocity difference across the shear layer increases the amplitude levels of the temperature disturbance at all Mach number. Also, the vertical extent of these disturbances seem to increase with increasing the velocity difference, especially towards the slow stream (for example, in figure \ref{f4}, the top boundary of the disturbance is roughly in $y=1$ for all velocity differences, while the bottom boundary is roughly in $y=0.9$ for $\Delta V=80\%$ and in $y=1.2$ for $\Delta V=50\%$).

\begin{figure}[htp]
 \begin{center}
  \includegraphics[width=4.2cm]{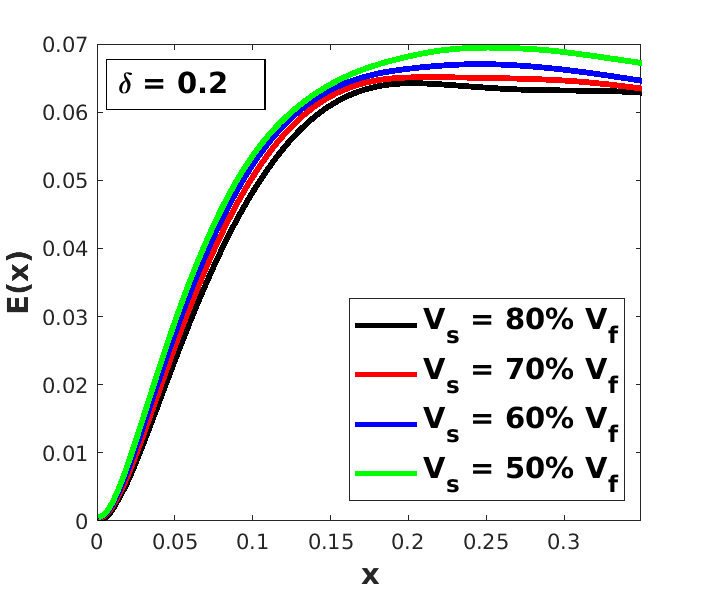}
  \includegraphics[width=4.2cm]{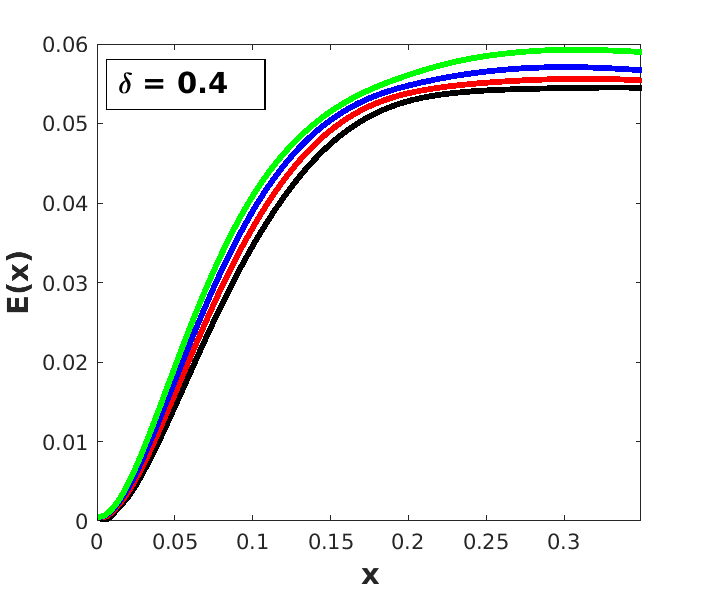}
  \includegraphics[width=4.2cm]{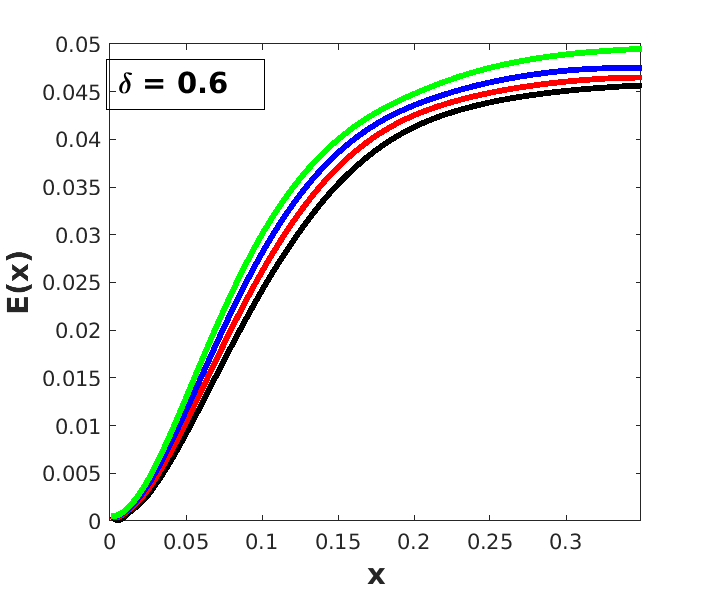}
 \end{center}
  \caption{Vortex energy distribution of different parametric settings for the $\boldsymbol{M=2}$ case.}
  \label{f7}
\end{figure}

\begin{figure}[htp]
 \begin{center}
  \includegraphics[width=4.2cm]{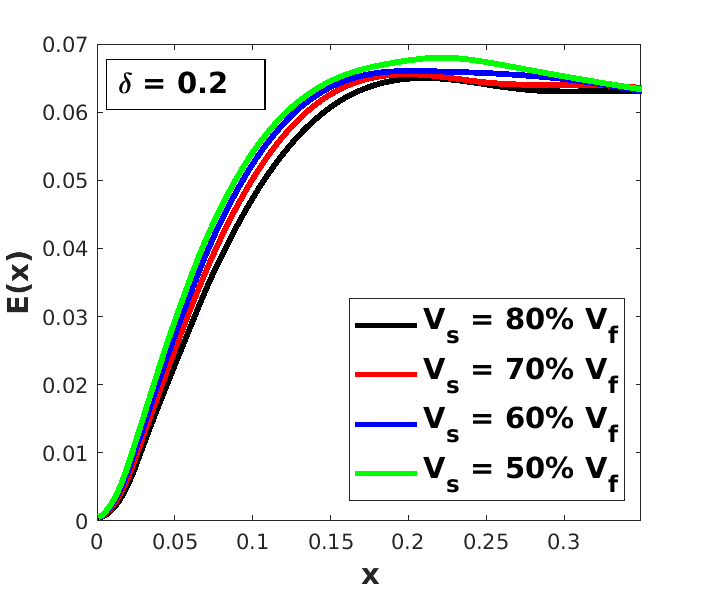}
  \includegraphics[width=4.2cm]{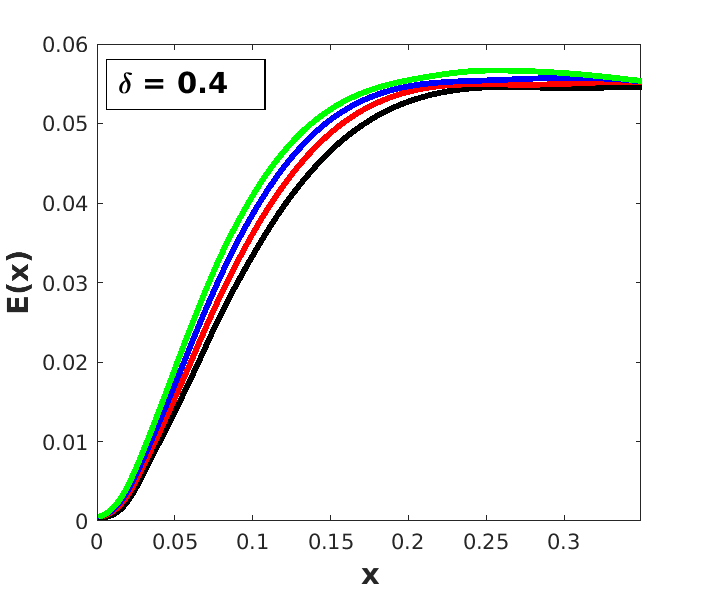} 
  \includegraphics[width=4.2cm]{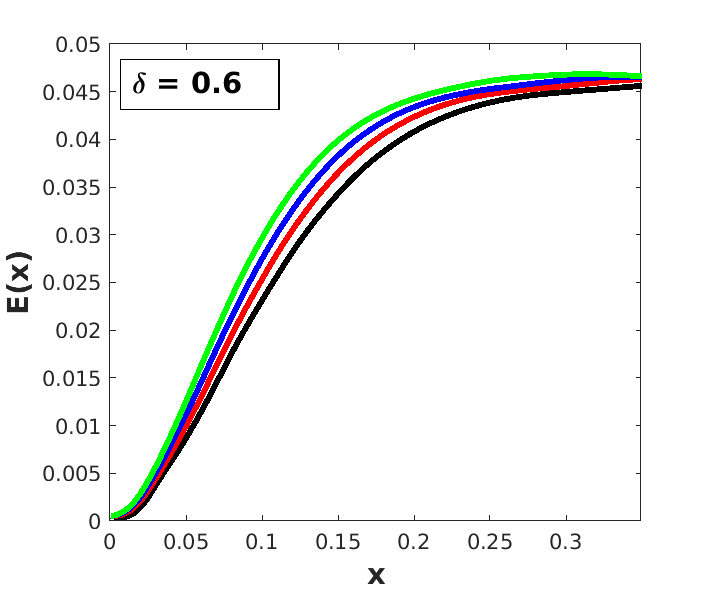} 
 \end{center}
  \caption{ Vortex energy distribution of different parametric settings for the $\boldsymbol{M=4}$ case.}
  \label{f8}
\end{figure}

\begin{figure}[htp]
 \begin{center}
  \includegraphics[width=4.2cm]{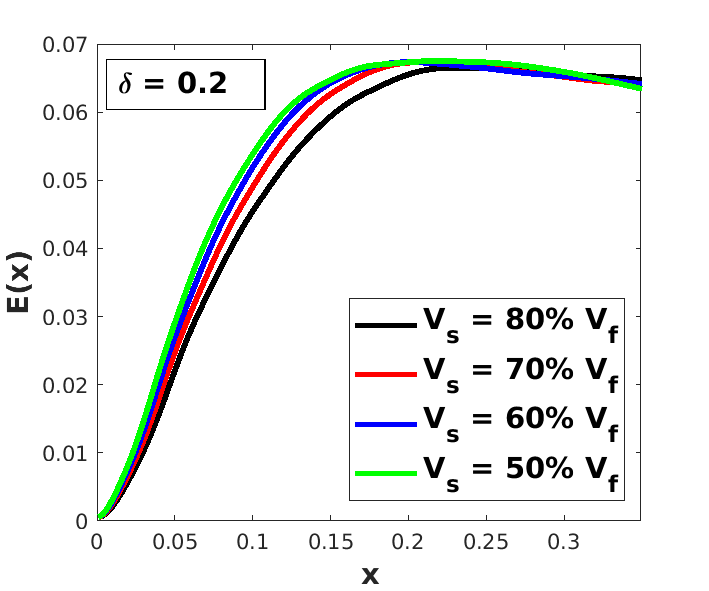}
  \includegraphics[width=4.2cm]{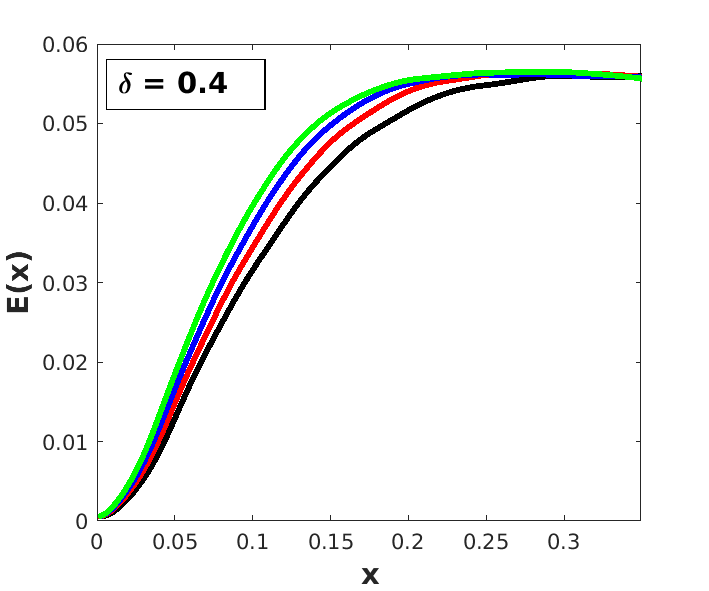} 
  \includegraphics[width=4.2cm]{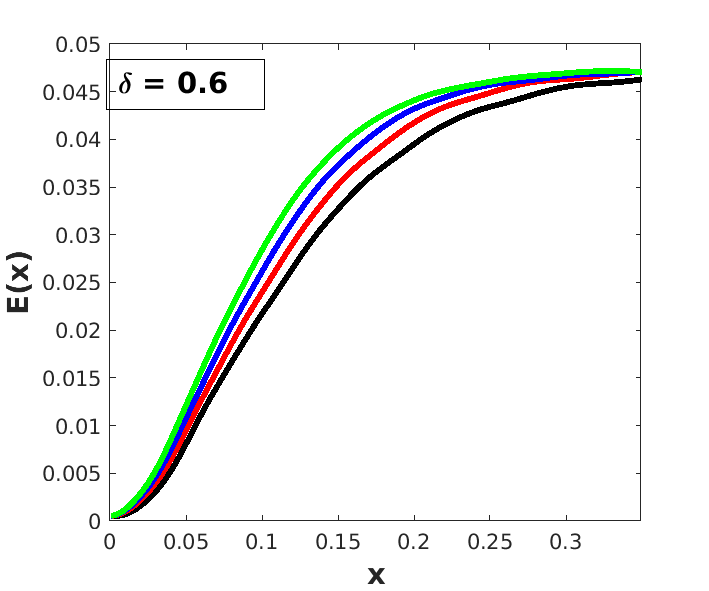} 
 \end{center}
  \caption{Vortex energy distribution of different parametric settings for the $\boldsymbol{M=6}$ case.}
  \label{f9}
\end{figure}

\begin{figure}[htp]
 \begin{center}
  \includegraphics[width=4.2cm]{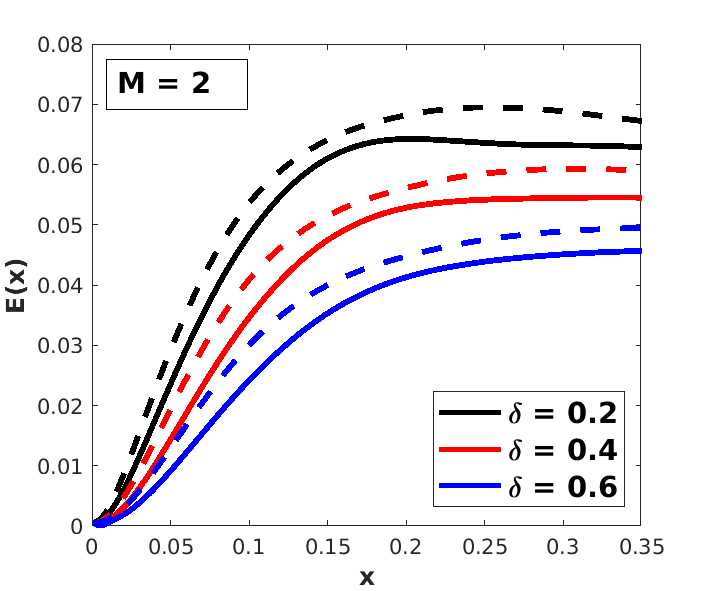}
  \includegraphics[width=4.2cm]{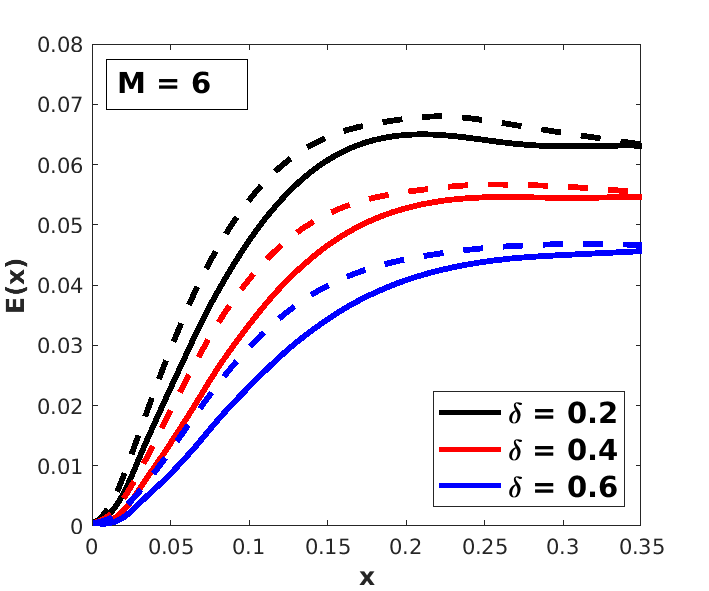}
  \includegraphics[width=4.2cm]{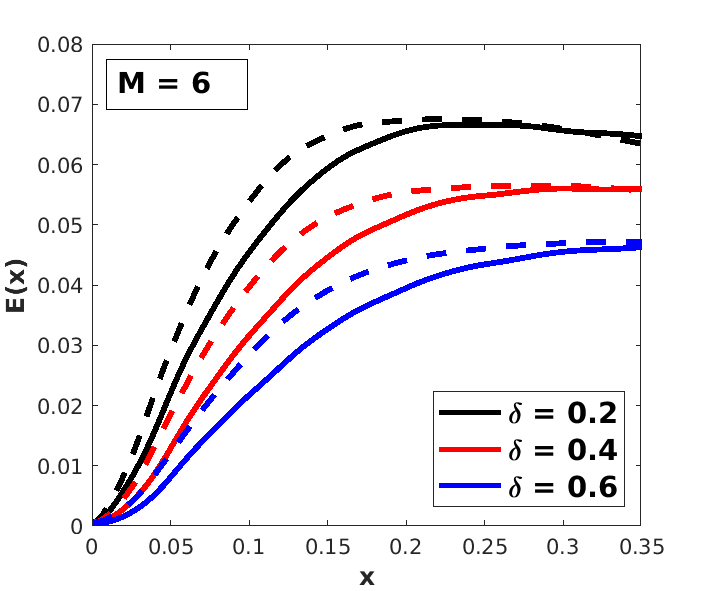} \\
 (a) \hspace{4.cm}(b) \hspace{4.cm}(c)
 \end{center}
  \caption{Effect of the shear layer thickness $\delta$ variation on the vortex energy distribution for $\Delta V = 80\%$ (solid lines) and $\Delta V = 50\%$ (dashed lines): a) $M=2$; b) $M=4$; c) $M=6$.}
  \label{f10}
\end{figure}

The kinetic energy distribution is calculated as 
\small
\begin{equation}\label{Energy_eq}
E(x) = \frac{1}{\Delta V^2} \intop_{z_1}^{z_2}  \intop_{0}^{\infty}  \left[ \left| u(x,y,z) - u_m(x,y) \right|^{2} +  \left| v(x,y,z) - v_m(x,y) \right|^{2} +  \left| w(x,y,z) - w_m(x,y) \right|^{2} \right] dzdy,
\end{equation}
\normalsize
where $u_m(x,y)$, $v_m(x,y)$, and $w_m(x,y)$ are the spanwise mean components of velocity, and $z_1$ and $z_2$ are the coordinates of the boundaries in the spanwise direction (note that the energy here is also scaled by $\Delta V^2$, which implies that the energy variation is the result of disturbances developing in the shear layer). In figures \ref{f7}-\ref{f9}, we plot $E(x)$ for different values of $\Delta V$ and in figure \ref{f10} we plot the same energy for different values of $\delta$. The disturbance energy $E(x)$ seems to be directly proportional to $\Delta V$ as it is highest for $\Delta V = 50\%$ and decreases as $\Delta V$ is reduced for all considered cases. Moreover, the streamwise location of the energy saturation (the point at which the energy starts to level off) moves farther downstream as $\Delta V$ decreases, especially for higher Mach numbers (see figure \ref{f9}). On the other hand, as seen in figure \ref{f10}, $E(x)$ is inversely proportional to $\delta$, which may be because viscous effects are more predominant in thicker shear layers; in this figure, we superposed the smallest (solid line) and the highest (dashed line) velocity difference levels, indicating that the the trend is the same for both (the other two velocity difference levels fall in between). We must point out that the energy reduction due to the increase in the shear layer thickness is rather significant, perhaps because the variation of the shear layer thickness is substantial (th e largest $\delta$ is three times greater that the smallest $\delta$). 

\section{Conclusions}

In this paper, we investigate the nonlinear development of centrifugal instabilities in a compressible curved free shear layer flow using a numerical solution to the boundary region equations, specifically, a parabolized form of the Navier-Stokes equations. Our focus lies on understanding the characteristics of these centrifugal instabilities, which exhibit similarities with the development of G"{o}rtler vortices in boundary layer flows over concave surfaces. The study encompasses variations in the free stream Mach number ($M_\infty$), the relative velocity difference between the two streams of the shear layer ($\Delta V$), the shear layer thickness ($\delta$), and the amplitude of the inflow disturbance ($A$).

Upon closer examination of the kinetic energy plots for the $M_\infty = 2$ case in figure \ref{f7}, we observe that $E(x)$ is directly proportional to $\Delta V$ and inversely proportional to $\delta$ across all the considered Mach numbers. However, the increase in shear layer thickness has an insignificant effect on energy reduction, with a mere $1\%$ drop in $E$ resulting from a $100\%$ increase in $\delta$, as evident in figure \ref{f10}. Furthermore, increasing the amplitude of the inflow disturbance ($A$) slightly boosts the kinetic energy, with less than $1\%$ increase in $E$ observed for a $100\%$ increase in $A$. Interestingly, a larger magnitude of $A$ hampers the influence of the relative velocity difference. The energy curves corresponding to different $\Delta V$ values exhibit a considerably reduced gap when comparing the $A = 0.02$ and $A = 0.04$ cases. Similar trends were observed in the parametric study of centrifugal instability development for the $M_\infty = 4$ and $M_\infty = 6$ cases.

Examining the contour plots of crossflow velocity magnitude in figures \ref{f1}-\ref{f3} for $\Delta V = 30\%$, we find that increasing the disturbance amplitude leads to significant growth in the mushroom-like structure's amplitude and renders the secondary structures more visible, indicating increased mixing for all Mach numbers under consideration. When comparing different Mach numbers at the same crossflow plane location, we observe a slower development of mushroom-like structures as $M_\infty$ increases, suggesting that achieving the same mixing efficiency would require traveling further downstream for higher Mach numbers. Consequently, numerical simulations for higher Mach number cases would incur higher computational costs due to the need for larger grid sizes to maintain comparable mixing efficiency.

\hspace{2mm}




\end{document}